\documentclass[11pt,aps,showpacs,nofootinbib,prd,aps,epsf,floats,axodraw,amsmath,amssymb,amsfonts]{revtex4}
\usepackage{amsmath, amssymb}
\usepackage{axodraw}
\newcommand{\mathsym}[1]{{}}

\usepackage{graphicx}
\usepackage{amsmath}
\usepackage{amssymb}
\usepackage{bm}
\newcommand{\be}{\begin{equation}}
\newcommand{\ee}{\end{equation}}
\newcommand{\bea}{\begin{eqnarray}}
\newcommand{\eea}{\end{eqnarray}}

\newcommand{\rem}[1]{}
\newsavebox{\PSLASH}
 \sbox{\PSLASH}{$p$\hspace{-1.8mm}/}
 
\renewcommand{\theequation}{\thesection.\arabic{equation}}
\newcounter{saveeqn}
\newcommand{\add}{\addtocounter{equation}{1}}
\newcommand{\alpheqn}{\setcounter{saveeqn}{\value{equation}}%
\setcounter{equation}{0}%
\renewcommand{\theequation}{\mbox{\thesection.\arabic{saveeqn}{\alph{equation}}}}}
\newcommand{\reseteqn}{\setcounter{equation}{\value{saveeqn}}%
\renewcommand{\theequation}{\thesection.\arabic{equation}}}

 \newsavebox{\notrightarrow}
 \sbox{\notrightarrow}{$\to$\hspace{-4mm}/}
 
 \newsavebox{\PARTIALSLASH}
 \sbox{\PARTIALSLASH}{$\partial$\hspace{-1.6mm}/}
 
 \newsavebox{\ASLASH}
 \sbox{\ASLASH}{$A$\hspace{-2.1mm}/}
 
 \newsavebox{\KSLASH}
 \sbox{\KSLASH}{$k$\hspace{-1.8mm}/}
 
 \newsavebox{\LSLASH}
 \sbox{\LSLASH}{$\ell$\hspace{-1.8mm}/}
 
 \newsavebox{\QSLASH}
 \sbox{\QSLASH}{$q$\hspace{-1.8mm}/}
 
 \newsavebox{\DSLASH}
 \sbox{\DSLASH}{$D$\hspace{-2.2mm}/}
 
 \newsavebox{\DbfSLASH}
 \sbox{\DbfSLASH}{${\mathbf D}$\hspace{-2.8mm}/}
 
 \newsavebox{\DELVECRIGHT}
 \sbox{\DELVECRIGHT}{$\stackrel{\rightarrow}{\partial}$}
 
 \newcommand{\blue}{\IfColor{\textCadetBlue}{}}
\newcommand{\black}{\IfColor{\textBlack}{}}
\newcommand{\red}{\IfColor{\textRed}{}}
\newcommand{\green}{\IfColor{\textOliveGreen}{}}
\newcommand{\lila}{\IfColor{\textRedViolet}{}}

\begin{document}
\title{On the mass spectrum of noncommutative Schwinger model in Euclidean $\mathbb{R}^{2}$ space}
\author{F. Ardalan$^{a,b}$}\email{ardalan@ipm.ir}
\author{M. Ghasemkhani$^{c}$}\email{ghasemkhani@ipm.ir}
\author{N. Sadooghi$^{a}$}\email{sadooghi@physics.sharif.ir}
\affiliation{$^a$Department of Physics, Sharif University of
Technology, P.O. Box 11155-9161, Tehran-Iran\\
$^{b}$Institute for Studies in Theoretical Physics and Mathematics (IPM)\\
School of Physics, P.O. Box 19395-5531, Tehran-Iran\\
$^{c}$Institute for Studies in Theoretical Physics and Mathematics (IPM)\\
School of Particles and Accelerators, P.O. Box 19395-5531,
Tehran-Iran}
\begin{abstract}
\noindent The mass spectrum of the noncommutative QED in
two-dimensional Euclidean $\mathbb{R}^{2}$ space is derived first in
a perturbative approach at one-loop level and then in a
nonperturbative approach using the equivalent bosonized
noncommutative effective action. It turns out that the mass spectrum
of noncommutative QED in two dimensions reduces to a single
non-interacting meson with mass $M_{\gamma}=\frac{g}{\sqrt{\pi}}$,
as in commutative Schwinger model.
\end{abstract}
\pacs{11.10.Nx, 11.10.Lm, 11.15.Bt} \maketitle
\section{Introduction}\label{Introduction}
\noindent Noncommutative field theories  \cite{noncommutative,
SWmap} are in general characterized by the replacement of familiar
product of functions with noncommutative Moyal star-product
\cite{moyal}, as the simplest realization of the canonical
commutation relation between space-time coordinates,
$[x_{\mu},x_{\nu}]_{\star}=i\theta_{\mu\nu}$. Their deformed
Lagrangian densities can therefore be written as an infinite series
in terms of higher order temporal and spatial derivative of
functions. From this point of view, noncommutative field theories
can be understood as a class of nonlocal higher-derivative field
theories, which appear in other areas of physics too. As it is
argued in \cite{simon}, unconstrained, nonlocal higher-derivative
theories, having more degrees of freedom than lower-derivative
theories, are dramatically different from their lower-derivative
counterparts. Nonlocal field theories appear, in general, as
effective field theories in a low-energy limit of a larger theory.
In particular, space-space noncommutative field theories describe
the low energy effective field theories of string theory in a
background magnetic field. In \cite{susskind}, space-time
noncommutative field theory is searched by considering open strings
in a constant background electric field. It is shown that here, in
contrast to the magnetic case, a critical electric field exists
beyond which the theory does not make sense, and that this critical
field prevents us from finding a limit in which the theory becomes a
field theory on a noncommutative space-time. In so far, string
theory does not admit a space-time noncommutative quantum field
theory, as its low energy limit, with the exception of light-like
noncommutativity \cite{barbon}. Whereas space-space noncommutative
field theories suffer from a mixing of ultraviolet and infrared
singularities in their perturbative dynamics \cite{minwalla},
space-time noncommutative theories in Minkowski space seem to be
seriously acausal and inconsistent with conventional Hamiltonian
evolution \cite{susskind}. Besides their $S$-matrix fail to be
unitary \cite{mehen}. In \cite{fredenhagen} attempts are made to
quantize theories with space-time noncommutativity, that leads,
however, to inconsitencies. A path integral formulation of
space-time noncommutativity using Schwinger's action principle is
introduced in \cite{fujikawa}, from which the canonical structure
for higher derivative theory is recovered. Using a simple field
theory non-local in time, it is shown that the quantization on the
basis of a naive interaction picture is not justified if the
interaction contains non-local terms in time. It is shown that a
unitary $S$-matrix can only be defined by using a modified time
ordering, but the positive energy condition is spoiled together with
a smooth Wick rotation to Euclidean theory. In \cite{ghasemkhani-1},
noncommutative QED in a 1+1 dimensional Minkowski space is expanded
in the order of the noncommutativity parameter $\theta$ up to
${\cal{O}}(\theta^{3})$. The resulting theory is non-local in time
and can be considered as a higher derivative theory including only a
finite number of time derivatives. Using the method of perturbative
quantization, it is then perturbatively quantized up to
${\cal{O}}(e^{2},\theta^{3})$, where $e$ is the corresponding QED
coupling constant. Recently, in \cite{tureanu}, space-time
noncommutative field theories are examined in various pictures and
quantization procedures. It is shown that different quantization
procedures lead to inconsistencies when time is taken as
noncommutative coordinate. The results in \cite{tureanu} are
consistent with the results in \cite{susskind,barbon,mehen,fujikawa}
and with string theory which, as we have mentioned before, ``does
not admit a space-time noncommutative quantum field theory, as its
low energy limit, with the exception of light-like
noncommutativity'' \cite{barbon}.
\par In this paper, we consider noncommutative QED in two-dimensional
\textit{Euclidean space} (noncommutative QED$_{2}$), where the
Minkowski time $x_{0}=t$ is replaced by the Euclidean imaginary time
$x_{2}\equiv it$, and the canonical commutation relation between the
coordinates is reduced to $[x_{i},x_{j}]=i\theta\epsilon_{ij}$, with
$i,j=1,2$ and $\epsilon_{ij}$ the two-dimensional antisymmetric
tensor. Hence, in contrast to its Minkowskian formulation, in the
Euclidean space the noncommutativity is defined between two
``equivalent'' spatial coordinates, and the noncommutative field
theory based on this formulation is therefore free of the before
mentioned problems of $1+1$ dimensional noncommutative field
theories.
\par
Having a Euclidean formulation of noncommutative QED$_{2}$, we are
in particular interested in the mass spectrum of the theory, that
plays the r$\hat{\mbox{o}}$le of the noncommutative counterpart of
the well-known Schwinger model \cite{schwinger}. Originally
introduced as a toy model to understand the confinement properties
of strong interaction, the Euclidean formulation of the Schwinger
model turns out to have many applications in system where the
physics is effectively reduced to two spatial dimensions. One
example of such a system is $3+1$ dimensional QED in the presence of
constant magnetic fields. For strong enough magnetic fields, aligned
in the third spatial direction, it is known that, because of the
phenomenon of magnetic catalysis \cite{catalysis}, the system is
reduced to a two-dimensional surface perpendicular to the direction
of the external magnetic field. The dynamics of electrons and
photons is then described effectively by a two-dimensional Euclidean
Schwinger model, where in a mechanism similar to the Higgs mechanism
the photons acquire a finite mass $m_{\gamma}\sim \frac{e}{L_{B}}$,
similar to the Schwinger's photon mass $M_{\gamma}\sim g$. Note that
whereas in $m_{\gamma}$, $e$ is the dimensionless coupling constant
of the $3+1$ dimensional QED, $g$ in $M_{\gamma}$ is the
dimensionful coupling constant of the two-dimensional QED. Moreover,
in $m_{\gamma}$, the magnetic length $L_{B}$ is defined by
$L_{B}\equiv |eB|^{-1/2}$. Remarkable, however, is the fact that the
effective two-dimensional field theory in the presence of strong
magnetic field can be effectively described by a deformed
noncommutative field theory in two spatial dimensions, where, for
$eB>0$, the commutation relation between the coordinates is given by
$[x_{i},x_{j}]=\kappa_{ij}$ with
$\kappa_{ij}=L_{B}^{2}\left(i\epsilon_{ij}-\delta_{ij}\right)$,
$i,j=1,2$ \cite{nc-magnet}. This turns out to be similar to the
corresponding relation in Euclidean noncommutative Schwinger model,
with the noncommutativity parameter $\theta\sim L_{B}^{2}$ playing
the r$\hat{\mbox{o}}$le of the squared of the magnetic length
$L_{B}$. What concerns the noncommutative version of the Euclidean
Schwinger model, we therefore expect that noncommutative photons
acquire a finite mass, that depends in a non-trivial way on the
dimensionful coupling constant $g$ and the noncommutativity
parameter $\theta$. It is the purpose of this paper to determine
these noncommutative corrections to the commutative photon mass
$M_{\gamma}$.
\par
Another point, which makes this problem far from trivial, is the
fact that noncommutative QED$_{2}$ has similar properties to
commutative QCD$_{2}$. This is because of the presence of
noncommutative star-products in the gauge kinetic term of the
Lagrangian density of the theory, including cubic and quartic
couplings of noncommutative gauge fields, as in commutative
QCD$_{2}$. Moreover, the equation of motion of noncommutative
QED$_{2}$ includes, similar to commutative QCD$_{2}$, a covariant
derivative of the noncommutative field strength tensor and is
therefore non-linear in the gauge field. Whereas in the commutative
Schwinger model, the photon mass $M_{\gamma}$ is determined
perturbatively from the pole of photon propagator, and turns out to
be one-loop exact \cite{schwinger}, the mass spectrum of QCD$_{2}$
is non-trivial and looking for it has a long history. The mass
spectrum of pure commutative two-dimensional single flavor
$U(N_{c})$ model was first determined by 't Hooft \cite{tHooft},
where in the $N_{c}\to \infty$ limit a nearly straight Regge
trajectory of confined mesonic states was found. Recently, the
spectrum of multi-flavor commutative $SU(N_{c})$ in two dimensions
is determined in \cite{armoni}. Here, the authors consider first a
(commutative) gauged bosonized action equivalent to QCD$_{2}$
action, that consists of a free Wess-Zumino-Witten (WZW) part, a
gauge kinetic part and an appropriate interaction part including the
WZW and gauge fields. Using then the light-cone momentum operator,
depending on the mesonic currents, they derive a 't Hooft-like mass
eigenvalue equation $P^{2}|\Phi\rangle=M^{2}|\Phi\rangle$, where
$|\Phi\rangle$ is the wave function of  ``currentball'' states of
QCD$_{2}$ in a Hilbert space which is truncated to two-current
states \cite{armoni}. Solving this eigenvalue equation numerically,
they arrive at the mass spectrum of QCD$_{2}$, that, in particular,
in the limit of large $N_{f}\gg N_{c}$, reduces to a single
noninteracting meson with mass $m=N_{f}M_{\gamma}$, i.e. $N_{f}$
copies of the commutative Schwinger mass
$M_{\gamma}=\frac{g}{\sqrt{\pi}}$. In the present paper, we will
closely follow the arguments of \cite{armoni}, and will determine
the mass spectrum of noncommutative QED$_{2}$ nonperturbatively. We
will arrive at the quite unexpected result that the mass-spectrum of
noncommutative QED$_{2}$ does not receive any noncommutative
correction and is therefore given by the spectrum of commutative
QED$_{2}$, consisting of a single free meson with mass
$M_{\gamma}=\frac{g}{\sqrt{\pi}}$. We will show that this result is
perturbatively exact in the noncommutativity parameter $\theta$, and
the noncommutative QED$_{2}$ coupling constant $g$. It is, however,
limited to the two current sector of the Hilbert space. This is in
contrast with the results arising in \cite{rahaman1}. Here, a
noncommutative $1+1$ dimensional bosonized Schwinger model is
considered. Using a perturbative Seiberg-Witten map \cite{SWmap} up
to first order in $\theta$, it is then mapped into an equivalent
gauge invariant commutative model. It is shown that the resulting
deformed theory consists of a massive boson, which is, in contrast
to its noncommutative counterpart, no longer free. The same
noncommutative bosonized Schwinger model, as in \cite{rahaman1}, is
also considered in \cite{rahaman2}, where, in particular, the
confinement-deconfienemt phase transition observed in commutative
Schwinger model is investigated. It is shown that though the
fuzziness of space-time introduces new features in the confinement
scenario, it does not affect the deconfining limit \cite{rahaman2}.
Schwinger model on fuzzy spheres is studied in \cite{harikumar}.
Other exactly solvable noncommutative models are discussed recently
in \cite{thirring} (see also the references therein).
\par
This paper is organized as follows: In Sec. II, after introducing
the noncommutative QED$_{2}$ model, we will determine the one-loop
correction to photon mass by determining the pole of the
corresponding photon propagator at one-loop level. In this order,
although additional diagrams relative to commutative QED$_{2}$ are
to be taken into account and they can potentially contribute to the
pole of the noncommutative photon propagator, the noncommutative
photon mass, being the same as the commutative photon mass, receives
no noncommutative corrections. To show that this result is one-loop
exact, we will follow, in Sec. III, the method used in \cite{armoni}
and determine the mass spectrum of noncommutative QED$_{2}$
nonperturbatively. In Sec. III.A, following the method introduced in
\cite{abdalla-2,abdalla-review} to bosonize commutative QCD$_{2}$,
we will first present the noncommutative Polyakov-Wiegmann fermionic
effective action and the full gauged bosonized action of
noncommutative QED$_{2}$, that includes, in particular, a
noncommutative WZW part. The latter coincides with the WZW action
previously determined in \cite{schaposnik} by explicitly computing
the fermion determinant of noncommutative QED$_{2}$. In Sec. III.B,
we will derive the energy-momentum tensor corresponding to the
gauged bosonized action of noncommutative QED$_{2}$, which will then
be used in Sec. III.C to determine the mass spectrum of
noncommutative QED$_{2}$ by solving the corresponding mass
eigenvalue equation. Section IV is devoted to our concluding
remarks.
\section{Photon mass in noncommutative QED$_{2}$: Perturbative approach}
\subsection{The model}
\setcounter{equation}{0}
\noindent  The Lagrangian density of noncommutative field theories
are given by their commutative counterpart with the commutative
product of functions replaced by noncommutative star-product,
defined by
\begin{eqnarray}\label{G1}
f(x)\star g(x)\equiv
\exp\left(\frac{i\theta_{\mu\nu}}{2}\frac{\partial}{\partial\xi_{\mu}}\frac{\partial}{\partial
\zeta_{\nu}}\right)\ f(x+\xi)g(x+\zeta)\Bigg|_{\xi=\zeta=0}.
\end{eqnarray}
Here, $\theta_{\mu\nu}$ is an antisymmetric constant matrix
reflecting the noncommutativity of space-time coordinates $x_{\mu}$
and $x_{\nu}$, $[x_{\mu},x_{\nu}]_{\star}\equiv i\theta_{\mu\nu}$.
In two-dimensional space-time coordinates $\theta_{01}$ and
$\theta_{10}=-\theta_{01}=-\theta$ are given in terms of
antisymmetric tensor of rank two $\epsilon_{\mu\nu}$ as
$\theta_{\mu\nu}=\theta\epsilon_{\mu\nu}$. In this paper,
noncommutative QED$_{2}$ will be considered in two-dimensional
Euclidean space, where $x^{0}=-ix^{2}$. The Lagrangian density of
two-dimensional noncommutative massless QED is then given by
\begin{eqnarray}\label{G2}
{\cal{L}}=-i\bar{\psi}\star\gamma_{\mu}\partial^{\mu} \psi+g
\bar{\psi}\star \gamma_{\mu}A^{\mu}\star
\psi+\frac{1}{4}F_{\mu\nu}\star
F^{\mu\nu}+\frac{1}{2\xi}(\partial_{\mu} A^{\mu})\star
(\partial_{\nu} A^{\nu})-\partial_{\mu}\bar c\star
(\partial^{\mu}c-ie[A_{\mu},c]_{\star}),\nonumber\\
\end{eqnarray}
with the Dirac $\gamma$-matrices $\gamma_{1}=\sigma_{2}$, and
$\gamma_{2}=-\sigma_{1}$, satisfying
$\{\gamma_{\mu},\gamma_{\nu}\}=2\delta_{\mu\nu}$ and
$[\gamma_{\mu},\gamma_{\nu}]=2i\epsilon_{\mu\nu}\gamma_{5}$, where
$\gamma_{5}=-i\gamma_{1}\gamma_{2}=\sigma_{3}$. Here, $\sigma_{i},
i=1,2,3$ are Pauli matrices. In (\ref{G2}), $\xi$ is the gauge
fixing parameter and the field strength tensor $F_{\mu\nu}$ is
defined by
\begin{eqnarray}\label{G3-a}
F_{\mu\nu}= \partial_{\mu} A_{\nu}-\partial_{\nu} A_{\mu}+ig
[A_{\mu},A_{\nu}]_{\star},
\end{eqnarray}
with $[A_{\mu},A_{\nu}]_{\star} = A_{\mu}\star A_{\nu}-A_{\nu}\star
A_{\mu}$. The Lagrangian density (\ref{G2}) is invariant under
\textit{global} $U(1)$ transformation
$\delta_{\alpha}\psi=i\alpha\psi$ and
$\delta_{\alpha}\bar\psi=-i\alpha\bar\psi$, which leads, according
to the prescriptions in \cite{sadooghi2001}, to two different
Noether currents
\begin{eqnarray}
J_{\mu}(x)&=&\psi(x)\star\bar{\psi}(x)\gamma_{\mu},\label{G4-a}\\
j_{\mu}(x)&=&\bar{\psi}(x)\gamma_{\mu}\star\psi(x).\label{G4-b}
\end{eqnarray}
Depending on their transformation properties under \textit{local}
$U(1)$ transformation, $\psi(x)\to e^{ig\alpha(x)}\star\psi(x)$, the
currents in (\ref{G4-a}) and (\ref{G4-b}) are denoted by covariant
and invariant currents, respectively. They satisfy the classical
continuity equations
\begin{eqnarray}\label{G5}
D_{\mu}J^{\mu}(x)=0,\qquad\mbox{and}\qquad
\partial_{\mu}j^{\mu}(x)=0,
\end{eqnarray}
with
$D_{\mu}J^{\mu}=\partial_{\mu}J^{\mu}-ig[A_{\mu},J^{\mu}]_{\star}$.
They arise from the equations of motion of massless fermionic fields
\begin{eqnarray}\label{G6}
\partial_{\mu}\bar{\psi}\gamma^{\mu}=ig\bar{\psi}\gamma^{\mu}\star
A_{\mu},\qquad\mbox{and}\qquad \gamma^{\mu}\partial_{\mu}\psi
=-igA_{\mu}\star\gamma^{\mu}\psi.
\end{eqnarray}
Similarly, there are two different axial vector currents
\begin{eqnarray}
J_{\mu}^{5}(x)&=&\psi(x)\star\bar{\psi}(x)\gamma_{\mu}\gamma_{5},
\label{G7}\\
j_{\mu}^{5}(x)&=&\bar{\psi}(x)\gamma_{\mu}\gamma_{5}\star\psi(x),
\label{G8}
\end{eqnarray}
associated with the global $U_{A}(1)$ axial invariance of the
Lagrangian density (\ref{G2}) under
$\delta_{\alpha}\psi=i\alpha\gamma_{5}\psi$ and
$\delta_{\alpha}\bar\psi=i\alpha\gamma_{5}\bar\psi$. They satisfy,
similar to the vector currents (\ref{G4-a}) and (\ref{G4-b}), two
different classical conservation laws
\begin{eqnarray}\label{G9}
D^{\mu}J_{\mu}^{5}(x)=0,\qquad\mbox{and}\qquad
\partial^{\mu}j_{\mu}^{5}(x)=0,
\end{eqnarray}
that can be derived also using the relation
$\gamma^{\mu}\gamma^{5}=-i\epsilon^{\mu\nu}\gamma_{\nu}$, which is
satisfied only in two dimensions, and the continuity relations
(\ref{G5}). In Sec. III, we will use the covariant axial vector
current (\ref{G7}), whose axial anomaly is given by
\cite{sadooghi2000}
\begin{eqnarray}\label{G10}
D^{\mu}J_{\mu}^{5}=-\frac{g}{2\pi}\epsilon_{\mu\nu}F^{\mu\nu},
\end{eqnarray}
to bosonize noncommutative QED$_{2}$.\footnote{For the axial anomaly
of the invariant axial vector current $j_{\mu}^{A}$ in two
dimensions see \cite{armoni-2002}. Nonplanar anomaly in $d=4$
dimensions was first determined in \cite{sadooghi2001}.}
\subsection{One-loop correction to the photon mass in noncommutative Schwinger model}
\noindent It is the purpose of this paper to determine possible
noncommutative corrections to the photon mass in two-dimensional
noncommutative Euclidean space. In this section, we will determine
the noncommutative photon mass perturbatively, by computing the pole
of noncommutative photon propagator at one-loop level. In ordinary
commutative Schwinger model, the photon mass is similarly determined
perturbatively from the pole of the full photon propagator
\begin{eqnarray}\label{G12}
{\cal{D}}^{\mu\nu}(p)=-\frac{1}{p^{2}[1+\Pi(p^{2})]}\left(\delta^{\mu\nu}-\frac{p^{\mu}p^{\nu}}{p^{2}}\right)-\xi~\frac{p^{\mu}p^{\nu}}{p^{4}},
\end{eqnarray}
where, $\Pi(p^{2})$ is the scalar function appearing in the photon
self-energy (the vacuum polarization tensor),
$\Pi^{\mu\nu}(p)=(p^{2}g^{\mu\nu}-p^{\mu}p^{\nu})\Pi(p^{2})$. At
one-loop level, $\Pi(p^{2})$ is given by dimensionally regularized
Feynman integral
\begin{eqnarray}\label{G13}
\Pi
(p^{2})=-\frac{2g^{2}}{(4\pi)^{\frac{d}{2}}}\int_{0}^{1}dx~x(1-x)\frac{\Gamma(2-\frac{d}{2})}{(-x(1-x)p^{2})^{2-\frac{d}{2}}},
\end{eqnarray}
corresponding to the one-loop correction to the photon propagator
from Fig. 1.
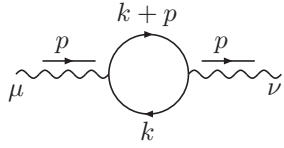
\begin{figure}[h] \SetScale{0.5}
  \begin{picture}(80,40) (0,0)
  \SetWidth{1.2}
    \Photon(50,0)(-20,0){3}{4}
    \ArrowArc(80,0)(-30,180,0)
    \ArrowArc(80,0)(-30,0,180)
    \Photon(110,0)(180,0){3}{4}
    \Text(-13,-10)[lb]{\small{\Black{{$\mu$}}}}
    \Text(85,-8)[lb]{\small{\Black{{$\nu$}}}}
    \Text(5,8)[lb]{{\Black{{$p$}}}}
    \Text(65,8)[lb]{{\Black{{$p$}}}}
    \Text(37,-25)[lb]{{\Black{{$k$}}}}
    \Text(28,18)[lb]{\small{\Black{{$k+p$}}}}
    \ArrowLine(120,10)(160,10)
    \ArrowLine(0,10)(40,10)
    \end{picture}
    \vspace{.5cm}
    \caption{Relevant one-loop photon self-energy diagram in commutative QED.}
\end{figure}
\par\noindent In $d=2$ dimensions, the Feynman integral (\ref{G13}) turns out to be finite and
yields $\Pi(p^{2})=\frac{g^{2}}{\pi p^{2}}$ with a pole in the limit
$p^{2}\to 0$. Plugging $\Pi(p^{2})$ in (\ref{G12}), the photon
propagator reads
\begin{eqnarray}\label{G14}
{\cal{D}}^{\mu\nu}(p)=-\frac{1}{p^{2}-\mu^{2}}\left(\delta^{\mu\nu}-\frac{p^{\mu}p^{\nu}}{p^{2}}\right)-\xi~\frac{p^{\mu}p^{\nu}}{p^{4}},
\end{eqnarray}
with the pole $\mu^{2}=\frac{g^{2}}{\pi}$, which can be interpreted
as the commutative photon mass $M_{\gamma}=\mu$, and turns out to be
one-loop exact \cite{schwinger}. Following the same method, we will
now determine the photon mass in two-dimensional noncommutative
Schwinger model at one-loop level. The general structure of photon
self-energy in $d$-dimensions is studied previously in \cite{ashok},
without referring directly to the photon mass in two-dimensions. In
the subsequent paragraph, we will present a brief review of these
results, and postpone the details of the computations to App. A.
\par
As it is argued in \cite{ashok}, in $d$-dimensions, the general
structure of the photon propagator is given by inverting the 1PI
two-point function and is given by
\begin{eqnarray}\label{G16}
D_{\mu\nu}=-\frac{1}{p^{2}+A}\left(\delta_{\mu\nu}-\frac{p_{\mu}p_{\nu}}{p^{2}}-\frac{\tilde{p}_{\mu}\tilde{p}_{\nu}}{\tilde{p}^{2}}
\right)-\frac{1}{p^{2}+A+B}\left(\frac{\tilde{p}_{\mu}\tilde{p}_{\nu}}{\tilde{p}^{2}}\right)
-\xi~\frac{p_{\mu}p_{\nu}}{p^{4}}.
\end{eqnarray}
Here, $\tilde{p}^{\mu}=\theta^{\mu\nu}p_{\nu}$. Moreover, $A$ and
$B$ are form factors arising from the vacuum polarization tensor
$\Pi^{\mu\nu}$
\begin{eqnarray}\label{G17}
\Pi^{\mu\nu}=A\left(\delta^{\mu\nu}-\frac{p^{\mu}p^{\nu}}{p^{2}}\right)+B\frac{\tilde{p}^{\mu}\tilde{p}^{\nu}}{\tilde{p}^{2}}.
\end{eqnarray}
Perturbatively, they receive contributions from Feynman integrals
corresponding to the vacuum polarization tensor $\Pi^{\mu\nu}$. They
are scalar functions in $p^{2}$ and $\tilde{p}$. In Fig. 2, the
relevant one-loop diagrams contributing to $A$ and $B$ in
noncommutative QED are shown.
\par
\begin{figure}[htb] \SetScale{0.5}
  \begin{picture}(80,40) (0,0)
  \SetWidth{1.2}
    \Photon(-300,0)(-370,0){3}{4}
    \ArrowArc(-270,0)(-30,180,0)
    \ArrowArc(-270,0)(-30,0,180)
    \Photon(-240,0)(-170,0){3}{4}
    \Text(-188,-10)[lb]{\small{\Black{{$\mu$}}}}
    \Text(-90,-8)[lb]{\small{\Black{{$\nu$}}}}
    \Text(-170,8)[lb]{{\Black{{$p$}}}}
    \Text(-110,8)[lb]{{\Black{{$p$}}}}
    \Text(-140,-25)[lb]{{\Black{{$k$}}}}
    \Text(-145,20)[lb]{\small{\Black{{$k+p$}}}}
    \ArrowLine(-230,7)(-190,7)
    \ArrowLine(-350,7)(-310,7)
    \Text(-144,-45)[lb]{(a)}
    \Text(-75,0)[lb]{+}
    \Photon(-105,0)(-35,0){3}{4}
    \PhotonArc(-5,0)(-30,180,0){4}{5}
    \PhotonArc(-5,0)(-30,0,180){4}{5}
    \Photon(25,0)(95,0){3}{4}
    \Text(-55,-10)[lb]{\small{\Black{{$\mu$}}}}
    \Text(43,-8)[lb]{\small{\Black{{$\nu$}}}}
    \Text(-37,8)[lb]{{\Black{{$p$}}}}
    \Text(28,8)[lb]{{\Black{{$p$}}}}
    \Text(-7,-25)[lb]{\small{\Black{{$k$}}}}
    \Text(-15,20)[lb]{\small{\Black{{$k+p$}}}}
    \ArrowLine(-85,7)(-45,7)
    \ArrowLine(45,7)(85,7)
    \ArrowArc(-5,0)(-15,135,-15)
    \Text(-10,-45)[lb]{(b)}
    \Text(58,0)[lb]{+}
    \Photon(160,0)(300,0){3}{6.5}
    \PhotonArc(230,36)(-30,180,0){3}{6}
    \PhotonArc(230,36)(-30,0,180){3}{6}
    \ArrowLine(160,7)(200,7)
    \ArrowLine(260,7)(300,7)
    \ArrowArc(230,36)(-15,135,-15)
    \Text(80,-10)[lb]{\small{$\mu$}}
    \Text(145,-8)[lb]{\small{{$\nu$}}}
    \Text(85,8)[lb]{{\Black{{$p$}}}}
    \Text(140,8)[lb]{{\Black{{$p$}}}}
    \Text(112,39)[lb]{\small{\Black{{$k$}}}}
    \Text(110,-45)[lb]{(c)}
    \Text(165,0)[lb]{+}
    \Photon(380,0)(440,0){3}{4}
    \CArc(470,0)(-30,0,15)
    \CArc(470,0)(-30,25,40)
    \CArc(470,0)(-30,50,65)
    \CArc(470,0)(-30,80,95)
    \CArc(470,0)(-30,105,120)
    \CArc(470,0)(-30,130,145)
    \CArc(470,0)(-30,155,170)
    \CArc(470,0)(-30,180,195)
    \CArc(470,0)(-30,205,220)
    \CArc(470,0)(-30,230,245)
    \CArc(470,0)(-30,255,270)
    \CArc(470,0)(-30,280,295)
    \CArc(470,0)(-30,305,320)
    \CArc(470,0)(-30,330,345)
    \CArc(470,0)(-30,355,360)
    \Photon(500,0)(560,0){3}{4}
    \ArrowLine(390,7)(420,7)
    \ArrowLine(520,7)(550,7)
    \ArrowArc(470,0)(-15,135,-15)
    \Text(190,-10)[lb]{\small{\Black{{$\mu$}}}}
    \Text(275,-8)[lb]{\small{\Black{{$\nu$}}}}
    \Text(205,8)[lb]{{\Black{{$p$}}}}
    \Text(270,8)[lb]{{\Black{{$p$}}}}
    \Text(232,-25)[lb]{\small{\Black{{$k$}}}}
    \Text(225,17)[lb]{\small{\Black{{$k+p$}}}}
    \Text(230,-45)[lb]{(d)}
    \end{picture}
    \vspace{1.5cm}
    \caption{Relevant one-loop photon self-energy diagrams in noncommutative QED.}
\end{figure}
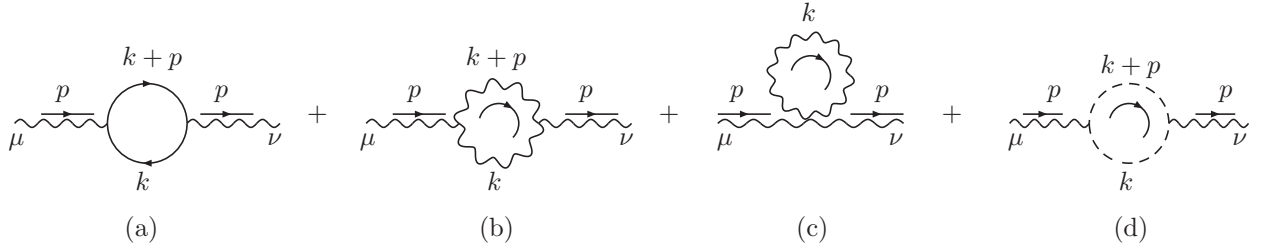
\par
Using the relation $\theta^{\mu\nu}=\theta\epsilon^{\mu\nu}$, which
is valid only in $d=2$ dimensions, it is easy to show that
$\tilde{p}_{\mu}\tilde{p}_{\nu}=\theta^{2}(p^{2}\delta_{\mu\nu}-p_{\mu}p_{\nu})$.
Plugging this relation in the vacuum polarization tensor (\ref{G17})
and the photon propagator $D^{\mu\nu}$ from (\ref{G16}), we get
\begin{eqnarray}\label{G18}
\Pi^{\mu\nu}=(A+B)\left(\delta^{\mu\nu}-\frac{p^{\mu}p^{\nu}}{p^{2}}\right),
\end{eqnarray}
as well as
\begin{eqnarray}\label{G19}
D_{\mu\nu}=-\frac{1}{p^{2}+(A+B)}\left(\delta_{\mu\nu}-\frac{p_{\mu}p_{\nu}}{p^{2}}\right)
-\xi~\frac{p_{\mu}p_{\nu}}{p^{4}}.
\end{eqnarray}
Identifying $A+B$ in the denominator of (\ref{G19}) with $A+B\equiv
p^{2}\Pi(p^{2})$, the noncommutative photon propagator (\ref{G19})
is comparable with (\ref{G12}) from commutative QED. Thus, according
to our description at the beginning of this section, in order to
determine the one-loop correction to the photon mass in
noncommutative field theory in two dimensions, it is enough to look
at $p^{2}\to 0$ behavior of $A+B$. Computing the corresponding
Feynman integrals to the diagrams of Fig. 2, we arrive after a
lengthy but straightforward calculation at (see App. A for more
details)
\begin{eqnarray}\label{G20}
A+B&=&-\mu^{2}-\frac{\mu^{2}}{4}\int_{0}^{1} dx~(x(1-x))^{-1}~
\bigg\{2(1+2x)(sK_{1}(s)-1)\nonumber\\
&&-(1-\xi)\big[6\left(2x^{2}-3x+1)(sK_{1}(s)-1\right)+\left(4x^{2}-6x+1\right)\left(s^{2}K_{2}(s)-2\right)\big]\nonumber\\
&&-\frac{1}{4}(1-\xi)^{2}(5s^{2}K_{2}(s)-2-s^{3}K_{3}(s)) \bigg\},
\end{eqnarray}
where $\mu^{2}\equiv \frac{g^{2}}{\pi}$, and $K_{n}(s)$ are the
modified Bessel-function of $n$-th order in terms of $s\equiv
p^{2}\theta\sqrt{x(1-x)}$. Keeping the noncommutativity parameter
$\theta$ finite and taking the IR limit $p^{2}\to 0$, or
equivalently $s\to 0$, we get
\begin{eqnarray}\label{G21}
 \lim_{p^{2}\rightarrow0}A+B=
 \lim_{p^{2}\rightarrow0}p^{2}\Pi(p^{2})=-\mu^{2}.
\end{eqnarray}
This is in particular, because of the relation
$s^{n}K_{n}(s)=2^{n-1}\Gamma[n]+{\cal{O}}(s^{2})$. Plugging
(\ref{G21}) in (\ref{G19}), it turns out that in noncommutative
Schwinger model the photon mass at one-loop level is still given by
\begin{eqnarray}\label{G22}
M_{\gamma}\equiv \mu=\frac{g}{\sqrt{\pi}},
\end{eqnarray}
and does not receive any correction proportional to the
noncommutative parameter $\theta$. The question of whether this
result is, as in the commutative case, exact in the orders of
$\theta$ and $g$ can not be answered within the framework of
perturbation theory. In the next section, we will bosonize the
theory; limiting ourselves to the two-current sector of the Hilbert
space and solving the corresponding mass eigenvalue equation, we
will show that the above result is indeed exact in noncommutative
parameter $\theta$ and coupling constant $g$.
\section{Photon mass  in noncommutative QED$_{2}$: Nonperturbative approach}
\setcounter{equation}{0}\noindent In the first part of this section,
we will briefly review the method leading to an appropriate
noncommutative bosonized action, which is equivalent to the
fermionic noncommutative action. Fot this purpose, we follow closely
the method introduced in \cite{abdalla-2} and \cite{abdalla-review}.
In the second part, the corresponding energy-momentum tensor for the
noncommutative bosonized gauged action, will be determined. Finally,
we will solve the mass eigenvalue equation of the theory and
determine the noncommutative photon mass nonperturbatively.
\subsection{Polyakov-Wiegmann functional and the gauged bosonized action of noncommutative Schwinger model}
\noindent In this section, we will derive the noncommutative
Polyakov-Wiegmann functional.\footnote{See \cite{polyakov,
abdalla-review, abdalla-2} for the commutative counterpart of this
functional in QCD$_{2}$.} Using the corresponding properties of this
functional under local vector and axial vector gauge transformation,
the gauged WZW action of noncommutative Schwinger model will be
derived, that, after inclusion of the gauge kinetic part of
noncommutative photons will eventually lead to the bosonized action
of noncommutative Schwinger model. To do this, we start, as in
commutative QCD$_{2}$, by choosing the gauge field $A_{\pm}$ in the
light-cone coordinates\footnote{The light-cone coordinates are
defined by $x_{\pm}=x_{1}\pm ix_{2}$. Similarly, $A_{\pm}=A_{1}\pm
iA_{2}$.} in terms of two fields $U$ and $V$ as
\begin{eqnarray}\label{H1}
A_{+}=\frac{i}{g}~U^{-1}\star\partial_{+}U, \qquad\mbox{and}\qquad
A_{-}=\frac{i}{g}~V\star\partial_{+}V^{-1},
\end{eqnarray}
where, $U$ and $V$ satisfy $U\star U^{-1}=1$ and $V\star V^{-1}=1$.
Moreover, under noncommutative $U(1)$ gauge transformation, $U$ and
$V$ transform as
\begin{eqnarray}\label{H2}
U\to U\star g_{v}^{-1},\qquad \mbox{and}\qquad V\to g_{v}\star V,
\end{eqnarray}
which is equivalent with the familiar $U(1)$ gauge transformation of
the gauge fields
\begin{eqnarray}\label{H3-a}
A_{\pm}\to g_{v}\star A_{\pm}\star g_{v}^{-1}+\frac{i}{g}g_{v}\star
\partial_{\pm}g_{v}^{-1},
\end{eqnarray}
with $A_{\pm}$ from (\ref{H1}), and $g_{v}\star g_{v}^{-1}=1$. To
determine $J_{\pm}$ in terms of $U$ and $V$, we use first the gauge
invariance of the fermionic determinant and choose, without loss of
generality, $V=1$ or equivalently $A_{-}=0$. Then, solving the
continuity equation (\ref{G5}) and the anomaly equation (\ref{G10})
in the light-cone coordinates and in $A_{-}=0$ gauge,
\begin{eqnarray}\label{H3}
\partial_{+}J_{-}+\partial_{-}J_{+}-ig[A_{+},
J_{-}]_{\star}&=&0,\nonumber\\
\partial_{-}J_{+}-\partial_{+}J_{-}+ig[A_{+},J_{-}]_{\star}&=&\frac{g}{\pi}\partial_{-}A_{+},
\end{eqnarray}
we arrive at
\begin{eqnarray}\label{H4}
J_{\pm}=\pm\frac{i}{2\pi}~U^{-1}\star\partial_{\pm}U.
\end{eqnarray}
The effective action $W[A]$ is then obtained using
$J_{-}=\frac{2}{g}\frac{\delta W[A]}{\delta A_{+}}$,
\begin{eqnarray}\label{H5}
\delta W[A]=\frac{g}{2}\int d^2 x ~J_{-}(x)\star\delta A_{+}(x).
\end{eqnarray}
Varying $A_{+}$ from (\ref{H1}), we get
\begin{eqnarray}\label{H6}
-ig\delta A_{+}=D_{+}(U^{-1}\star\delta U),\qquad\mbox{where}\qquad
D_{+}f\equiv
\partial_{+}f+[U^{-1}\star\partial_{+}U,f]_{\star}.
\end{eqnarray}
Plugging $\delta A_{+}$ from (\ref{H6}) together with $J_{-}$ from
(\ref{H4}) in (\ref{H5}), and using, after integrating (\ref{H5}) by
part, the identity $D_{-}(U^{-1}\star\partial_{+}
U)=\partial_{+}(U^{-1}\star\partial_{-} U)$,  $\delta W$ can be
rewritten as
\begin{eqnarray}\label{H7}
\delta W=-\frac{1}{4\pi}~\int d^{2}x~(U^{-1}\star\delta
U)\star\partial_{-}(U^{-1}\star\partial_{+}U).
\end{eqnarray}
This equation may then be integrated in terms of the WZW effective
action, $\Gamma[U]=-W[A]$, which consists of two parts,
$\Gamma[U]=S_{\mbox{\tiny{P$\sigma$M}}}[U]+S_{\mbox{\tiny{WZ}}}[U]$.
Here, $S_{\mbox{\tiny{P$\sigma$M}}}[U]$ is the action of the
principal $\sigma$-model, which is given by
\begin{eqnarray}\label{H8}
S_{\mbox{\tiny{P$\sigma$M}}}[U]=\frac{1}{8\pi}~\int
d^{2}x~(\partial_{\mu}U^{-1})\star(\partial^{\mu}U)=-\frac{1}{8\pi}\int
d^{2}x\left(U\star\partial_{\mu}U^{-1}\right)\star\left(U\star\partial^{\mu}U^{-1}\right),
\end{eqnarray}
and $S_{\mbox{\tiny{WZ}}}[U]$ is the Wess-Zumino term, that reads
\begin{eqnarray}\label{H9}
S_{\mbox{\tiny{WZ}}}[U]=-\frac{i}{4\pi}~\int_{0}^{1}dr\int
d^{2}x~\epsilon^{\mu\nu}U^{-1}_{r}\star\dot{U}_{r}\star
U^{-1}_{r}\star\partial_{\mu}U_{r}\star
U^{-1}_{r}\star\partial_{\nu}U_{r}.
\end{eqnarray}
In (\ref{H9}), $U_{r}$ is the extension of $U$ and satisfies the
boundary conditions $U_{r}(x)\big|_{r=1}=U$, and
$U_{r}(x)\big|_{r=0}=1$. Moreover, $\dot{U}_{r}\equiv
\partial_{r}U_{r}$. To obtain the noncommutative Polyakov-Wiegmann
functional, the effective action $\Gamma[U]$ is to be redefined in
terms of gauge invariant combination $\Sigma\equiv U\star V$. We
arrive first at
$\Gamma[\Sigma]=S_{\mbox{\tiny{P$\sigma$M}}}[\Sigma]+S_{\mbox{\tiny{WZ}}}[\Sigma]$,
with the principal $\sigma$-model part from (\ref{H8}),
\begin{eqnarray}\label{H11}
S_{\mbox{\tiny{P$\sigma$M}}}[\Sigma]&=&
S_{\mbox{\tiny{P$\sigma$M}}}[U]+S_{\mbox{\tiny{P$\sigma$M}}}[V]+\frac{1}{4\pi}\int
d^{2}x~(U^{-1}\star\partial_{\mu}U)\star(V\star\partial^{\mu}V^{-1}),
\end{eqnarray}
and the Wess-Zumino part from (\ref{H9})
\begin{eqnarray}\label{H12}
S_{\mbox{\tiny{WZ}}}[\Sigma]=S_{\mbox{\tiny{WZ}}}[U]+S_{\mbox{\tiny{WZ}}}[V]-\frac{i}{4\pi}\int_{0}^{1}dr\int
d^{2}x~\epsilon^{\mu\nu}{\cal W}_{\mu\nu},
\end{eqnarray}
with
\begin{eqnarray}\label{H13}
{\cal
W}_{\mu\nu}=\frac{d}{dr}\left(U^{-1}_{r}\star\partial_{\mu}U_{r}\star
V_{r}\star\partial_{\nu}V^{-1}_{r}\right)-
\partial_{\mu}\left(V_{r}\star\partial_{\nu}V^{-1}_{r}\star
U^{-1}_{r}\star\dot{U}_{r}\right)-\partial_{\nu}\left(U^{-1}_{r}\star\partial_{\mu}U_{r}\star
V_{r}\star\dot{V}_{r}^{-1}\right).\nonumber\\
\end{eqnarray}
Neglecting then the last two terms in (\ref{H13}), that yield two
vanishing surface integrals after integrating over the coordinates
in (\ref{H12}), and using the boundary conditions for $U_{r}$ and
$V_{r}$,\footnote{Here, it is assumed that $V_{r}$ satisfies the
same boundary conditions as $U_{r}$.} we arrive at the
noncommutative Polyakov-Wiegmann functional for $\Gamma[\Sigma]$,
\begin{eqnarray}\label{H14}
\Gamma[\Sigma]=\Gamma[U]+\Gamma[V]+\frac{1}{4\pi}\int
d^{2}x~\left(\delta^{\mu\nu}-i\epsilon^{\mu\nu}\right)\left(U^{-1}\star\partial_{\mu}U\right)\star\left(V\star\partial_{\nu}V^{-1}\right),
\end{eqnarray}
or equivalently in the light-cone coordinates, at
\begin{eqnarray}\label{H15}
\Gamma[\Sigma]=\Gamma[U]+\Gamma[V]+\frac{1}{4\pi}\int
d^{2}x~\left(U^{-1}\star\partial_{+}U\right)\star\left(V\star\partial_{-}V^{-1}\right).
\end{eqnarray}
Note that to derive the factorized forms (\ref{H11}) and (\ref{H12})
of $S_{\mbox{\tiny{P$\sigma$M}}}[\Sigma]$ as well as
$S_{\mbox{\tiny{WZ}}}[\Sigma]$, extensive use is made of the trace
property of the star-product of two functions under a
two-dimensional integral $\int d^{2}x~f\star g=\int d^{2}x~g\star
f$, and the relations $\partial_{\mu}U\star U^{-1}=-U\star
\partial_{\mu}U^{-1}$, as well as $\partial_{\mu}V\star V^{-1}=-V\star
\partial_{\mu}V^{-1}$, arising from $U\star U^{-1}=1$ and $V\star
V^{-1}=1$. Although the Polyakov-Wiegmann functional (\ref{H14}) is
invariant under gauge transformation (\ref{H2}), it is not invariant
under the $U_{A}(1)$ axial gauge transformation
\begin{eqnarray}\label{H16}
U\to U\star g_{a}^{-1},\qquad \mbox{and}\qquad V\to g^{-1}_{a}\star
V,
\end{eqnarray}
or equivalently
\begin{eqnarray}\label{H17}
A_{+}&\to& g_{a}\star A_{+}\star
g_{a}^{-1}+\frac{i}{g}g_{a}\star\partial_{+}g_{a}^{-1},\nonumber\\
A_{-}&\to& g_{a}^{-1}\star A_{-}\star
g_{a}+\frac{i}{g}g_{a}^{-1}\star\partial_{-}g_{a},
\end{eqnarray}
with $g_{a}\star g_{a}^{-1}=1$. This non-invariance can be expressed
in an equivalent gauged bosonic action for the fermions, defined by
\begin{eqnarray}\label{H18}
S_F[A,\omega]\equiv\Gamma[\Sigma;\omega]-\Gamma[\Sigma],
\end{eqnarray}
Here, the bosonized action $S_{F}[A,\omega]$  remains invariant
under local gauge transformation (\ref{H2}) of $U$ and $V$ fields
and the vector gauge transformation of the WZW field $\omega$, i.e.
$\omega\to g_{v}\star \omega\star g_{v}^{-1}$ \cite{abdalla-2}. The
explicit form of $S[A,\omega]$ can be found by successive use of the
Polyakov-Wiegmann functional in the light-cone coordinates
(\ref{H15}), and is given by
\begin{eqnarray}\label{H19}
S_{F}[A,\omega]=\Gamma[\omega]+\frac{1}{4\pi}\int
d^{2}x~{\cal{W}}[A,\omega],
\end{eqnarray}
with ${\cal{W}}[A,\omega]$ defined by
\begin{eqnarray}\label{H20}
{\cal{W}}[A,\omega]\equiv g^{2}A_{+}\star
A_{-}-g^{2}A_{+}\star\omega\star A_{-}\star \omega^{-1}-igA_{+}\star
\omega\star\partial_{-}\omega^{-1}-igA_{-}\star\omega^{-1}\star\partial_{+}\omega,
\end{eqnarray}
where the definitions of $A_{\pm}$ in terms of $U$ and $V$ from
(\ref{H1}) are used. Using ${\cal{J}}_{\pm}\equiv
\frac{2}{g}\frac{\delta S_{F}[A,\omega]}{\delta A_{\mp}}$, with
$S_{F}[A,\omega]$ from (\ref{H19})-(\ref{H20}), the corresponding
currents are given by
\begin{eqnarray}\label{H21-a}
{\cal{J}}_{+}&=&\frac{1}{2\pi}\left(gA_{+}-g\omega^{-1}\star
A_{+}\star\omega-i\omega^{-1}\star
\partial_{+}\omega\right),\nonumber\\
{\cal{J}}_{-}&=&\frac{1}{2\pi}\left(gA_{-}-g\omega\star
A_{-}\star\omega^{-1}-i\omega\star
\partial_{-}\omega^{-1}\right).
\end{eqnarray}
Note that in light-cone gauge $A_{-}=0$, ${\cal{J}}_{-}$ from
(\ref{H21-a}) is reduced to
$${\cal{J}}_{-}=J_{-}[\omega]=-\frac{i}{2\pi}\omega\star\partial_{-}\omega^{-1}$$
which  is of the same form as the currents $J_{-}$ from (\ref{H4})
of the free theory, rewritten in terms of $\omega$. Same result
arises also in commutative QCD$_{2}$ in the light-cone gauge
\cite{witten}.\footnote{Similarly one can show that since in the
light-cone gauge $S_{F}[A,\omega]$ from (\ref{H19}) reduces to
$S_{F}[A,\omega]=\Gamma[\omega]-\frac{ig}{4\pi}\int
d^{2}x~A_{+}\star\omega\star\partial_{-}\omega^{-1}$, the only
current coupled to $A_{+}$ is ${\cal{J}}_{-}=\frac{2}{g}\frac{\delta
S_{F}}{\delta A_{+}}=-\frac{i}{2\pi}\omega\star\partial_{-}\omega
=J_{-}[\omega]$.} Moreover, it can be shown that since $J_{\pm}$
satisfy the equations
\begin{eqnarray}\label{H21-c}
\partial_{\mp}J_{\pm}=0,
\end{eqnarray}
arising from the variation of $\Gamma[\omega]$, ${\cal{J}}_{-}$ in
light-cone gauge is, as in QCD$_{2}$, only a function of $x^{-}$.
Later, we will use this fact to show that the spectrum of
noncommutative QED$_{2}$ reduces to the spectrum of commutative
Schwinger model. Adding, at this stage, the gauge kinetic action
$\sim \int d^{2}x F^{\mu\nu}F_{\mu\nu}$ to (\ref{H19}), we arrive at
the full gauged bosonized action of noncommutative QED$_{2}$
\begin{eqnarray}\label{H21}
\lefteqn{S_{b}[A,\omega]=\Gamma[\omega]+\frac{1}{4}\int
d^{2}x~F_{\mu\nu}\star F^{\mu\nu}}\nonumber\\
&&+\frac{1}{4\pi}\int d^{2}x~\bigg\{g^{2}A_{\mu}\star
A^{\mu}-g^{2}\kappa_{-}^{\mu\nu}A_{\mu}\star\omega\star
A_{\nu}\star\omega^{-1}-ig\kappa_{-}^{\mu\nu}A_{\mu}\star\omega\star\partial_{\nu}\omega^{-1}
-ig
\kappa_{+}^{\mu\nu}A_{\mu}\star\omega^{-1}\star\partial_{\nu}\omega\bigg\},\nonumber\\
\end{eqnarray}
where $\kappa_{\pm}^{\mu\nu}\equiv \delta^{\mu\nu}\pm
i\epsilon^{\mu\nu}$. In the light-cone gauge $A_{-}=0$, the
bosonized action (\ref{H21}) is further reduced to
\begin{eqnarray}\label{H22}
S_{b}[\omega]=\Gamma[\omega]-\frac{g^{2}}{2}\int d^{2}x~J_{-}\star
\frac{1}{\partial_{-}^{2}}J_{-}.
\end{eqnarray}
Here, $J_{-}=-\frac{i}{2\pi}\omega\star\partial_{-}\omega^{-1}$. To
arrive at (\ref{H22}), we have replaced $F_{\mu\nu}F^{\mu\nu}$ in
(\ref{H21}) with
$F_{\mu\nu}F^{\mu\nu}=-\frac{1}{2}(\partial_{-}A_{+})^{2}$ and used
the equation motion of the gauge fields
$D_{\mu}F^{\mu\nu}=-gJ^{\nu}$, that in the light-cone gauge reduces
to $\partial_{-}^{2}A_{+}=2gJ_{-}$. In the next section, the
energy-momentum tensor corresponding to (\ref{H21}) will be derived.
\subsection{Energy-momentum tensor}
\noindent To determine the mass spectrum of noncommutative Schwinger
model, we have to solve the eigenvalue equation
$P^{2}|\Phi\rangle=M_{}^{2}|\Phi\rangle$, that in light-cone space
reduces to $P^{+}P^{-}|\Phi\rangle=M_{}^{2}|\Phi\rangle$, with
$P^{\pm}\sim \int dx^{-}T^{+\pm}$. Here, $T^{+\pm}$ are the
components of the total energy-momentum tensor $T^{\mu\nu}$
corresponding to the full bosonized action (\ref{H21}) in the
light-cone coordinates. In this section, we use the general
definition of $T^{\mu\nu}$
\begin{eqnarray}\label{H24}
T^{\mu\nu}=\frac{2}{\sqrt{-g}}\frac{\delta S}{\delta
g_{\mu\nu}(x)}\bigg|_{g_{\mu\nu}=\delta_{\mu\nu}},
\end{eqnarray}
to determine the total energy momentum tensor
\begin{eqnarray}\label{H23}
T^{\mu\nu}=T^{\mu\nu}_{\mbox{\tiny{P$\sigma$M}}}+T^{\mu\nu}_{\mbox{\tiny{WZ}}}+T^{\mu\nu}_{\mbox{\tiny{gauge}}}+T^{\mu\nu}_{\mbox{\tiny{${\cal{W}}$}}}.
\end{eqnarray}
Plugging first the action of the principal $\sigma$-model
$S_{\mbox{\tiny{P$\sigma$M}}}[\omega]$ from (\ref{H8}) in
(\ref{H24}), the corresponding energy momentum-tensor
$T_{\mbox{\tiny{P$\sigma$M}}}^{\mu\nu}$ in a Sugawara form
\cite{sugawara} reads
\begin{eqnarray}\label{H25}
T_{\mbox{\tiny{P$\sigma$M}}}^{\mu\nu}=\frac{\pi}{2}\bigg(J^{\mu}(x)\star
J^{\nu}(x)+J^{\nu}(x)\star
J^{\mu}(x)-\delta^{\mu\nu}J_{\lambda}(x)\star J^{\lambda}(x)\bigg).
\end{eqnarray}
In the light-cone gauge $A_{-}=0$, where
$J^{+}=J_{-}=-\frac{i}{2\pi}\omega\star\partial_{-}\omega^{-1}$, the
components of $T_{\mbox{\tiny{P$\sigma$M}}}^{\mu\nu}$ are given by
\begin{eqnarray}\label{H26}
T_{\mbox{\tiny{P$\sigma$M}}}^{++}=\pi J^{+}\star
J^{+},\qquad\mbox{and}\qquad
T_{\mbox{\tiny{P$\sigma$M}}}^{+-}=T_{\mbox{\tiny{P$\sigma$M}}}^{--}=T_{\mbox{\tiny{P$\sigma$M}}}^{-+}=0.
\end{eqnarray}
Similarly, the energy-momentum tensor
$T^{\mu\nu}_{\mbox{\tiny{WZ}}}$, corresponding to the Wess-Zumino
part of the effective action $S_{\mbox{\tiny{WZ}}}[\omega]$ from
(\ref{H9}) is determined using (\ref{H24}) and reads
\begin{eqnarray}\label{H27}
\lefteqn{T_{\mbox{\tiny{WZ}}}^{\mu\nu}=-\frac{i}{4\pi}\int_{0}^{1}dr\int
d^{2}x~
\bigg(-\delta^{\mu\nu}\epsilon^{\rho\sigma}\omega_{r}^{-1}\star\dot\omega_{r}\star\omega_{r}^{-1}\star\partial_{\rho}\omega_{r}\star
\omega_{r}^{-1}\star\partial_{\sigma}\omega_{r}}\nonumber\\
&&+
\epsilon^{\mu\sigma}\omega_{r}^{-1}\star\dot\omega_{r}\star\omega_{r}^{-1}\star\partial^{\nu}\omega_{r}\star\omega_{r}^{-1}\star\partial_{\sigma}\omega_{r}+
\epsilon^{\sigma\mu}\omega_{r}^{-1}\star\dot\omega_{r}\star\omega_{r}^{-1}\star\partial_{\sigma}\omega_{r}\star\omega_{r}^{-1}\star\partial^{\nu}\omega_{r}
\bigg).
\end{eqnarray}
As it turns out, in the light-cone coordinates, all the components
of the WZ part of the energy-momentum tensor vanish, in particluar
$T_{\mbox{\tiny{WZ}}}^{++}=T_{\mbox{\tiny{WZ}}}^{+-}=0$. As for the
noncommutative energy-momentum tensor corresponding to the gauge
kinetic action $\sim\int d^{2}x F_{\mu\nu}^2$ from (\ref{H21}), it
is previously determined in \cite{das-2}, using the (\ref{H24}). It
is given by
\begin{eqnarray}\label{H28}
T_{\mbox{\tiny{gauge}}}^{\mu\nu}=\frac{1}{2}\left(F^{\nu\alpha}\star
F^{\mu}_{\alpha}+F^{\mu\alpha}\star
F^{\nu}_{\alpha}-\frac{1}{2}\delta^{\mu\nu}F^{\alpha\beta}\star
F_{\alpha\beta}\right),
\end{eqnarray}
and reduces in the light-cone gauge to
\begin{eqnarray}\label{H29}
T_{\mbox{\tiny{gauge}}}^{++}=T_{\mbox{\tiny{gauge}}}^{--}=0,&\qquad\qquad&
T_{\mbox{\tiny{gauge}}}^{+-}=T_{\mbox{\tiny{gauge}}}^{-+}=\frac{1}{8}(\partial^{+}A^{-})\star(\partial^{+}A^{-}).
\end{eqnarray}
Finally, the corresponding energy-momentum tensor to ${\cal{W}}$
from (\ref{H20}) [or in $\mu,\nu$ coordinates from the last term of
(\ref{H21})] is given by
\begin{eqnarray}\label{H30}
T^{\mu\nu}_{\mbox{\tiny{${\cal{W}}$}}}&=&\frac{1}{4\pi}
\bigg\{\delta^{\mu\nu}{\cal{W}}(x)-g^{2}(A^{\mu}\star
A^{\nu}+A^{\nu}\star A^{\mu})+g^{2}(A^{\mu}\star\omega\star
A^{\nu}\star\omega^{-1}+A^{\nu}\star\omega\star
A^{\mu}\star\omega^{-1})\nonumber\\
&&-ig^{2}(\epsilon^{\mu\sigma}A^{\nu}\star\omega\star
A_{\sigma}\star\omega^{-1}+\epsilon^{\sigma\mu}A_{\sigma}\star\omega\star
A^{\nu}\star\omega^{-1})+ig(A^{\mu}\star\omega\star\partial^{\nu}\omega^{-1}+A^{\nu}\star\omega\star\partial^{\mu}\omega^{-1})\nonumber\\
&&+g(\epsilon^{\mu\sigma}A^{\nu}\star\omega\star\partial_{\sigma}\omega^{-1}+\epsilon^{\sigma\mu}A_{\sigma}\star\omega\star\partial^{\nu}\omega^{-1})
+ig(A^{\mu}\star\omega^{-1}\star\partial^{\nu}\omega+A^{\nu}\star\omega^{-1}\star\partial^{\mu}\omega)\nonumber\\
&&-g(\epsilon^{\mu\sigma}A^{\nu}\star\omega^{-1}\star\partial_{\sigma}\omega+\epsilon^{\sigma\mu}A_{\sigma}\star\omega^{-1}\star\partial^{\nu}\omega)
\bigg\}.
\end{eqnarray}
It can be checked that in the light-cone gauge all the components of
$T^{\mu\nu}_{\mbox{\tiny{${\cal{W}}$}}}$ vanish, in particular
$T_{\mbox{\tiny{${\cal{W}}$}}}^{++}=T^{+-}_{\mbox{\tiny{${\cal{W}}$}}}=0$.
Adding all the contributions from (\ref{H25}), (\ref{H27}),
(\ref{H28}) and (\ref{H30}) to the relevant component of the total
energy-momentum tensor (\ref{H24}), we arrive at
\begin{eqnarray}\label{H31}
T^{++}=T^{++}_{\mbox{\tiny{P$\sigma$M}}}=\bar{J}\star \bar{J},
\qquad\mbox{and}\qquad T^{+-}=
T_{\mbox{\tiny{gauge}}}^{+-}=\frac{1}{8}(\partial^{+}A^{-})\star(\partial^{+}A^{-}),
\end{eqnarray}
where the identity $J^{+}=J_{-}$ and the redefinition $\bar{J}\equiv
\sqrt{\pi}J_{-}$ is used. Using (\ref{H31}), $P^{\pm}$ are defined
by
\begin{eqnarray}\label{H32}
P^{+}&=&\frac{1}{2}\int dx^{-}~T^{++}=\frac{1}{2}\int
dx^{-}\bar{J}(x^{-})\star\bar{J}(x^{-})=\frac{1}{2}\int
dx^{-}\bar{J}(x^{-})\bar{J}(x^{-}),\nonumber\\
P^{-}&=&\frac{1}{2}\int dx^{-}T^{+-}=-\frac{g^{2}}{4\pi}\int
dx^{-}\bar{J}(x^{-})\star\frac{1}{\partial_{-}^{2}}\bar{J}(x^{-})=-\frac{g^{2}}{4\pi}\int
dx^{-}\bar{J}(x^{-})\frac{1}{\partial_{-}^{2}}\bar{J}(x^{-}).
\end{eqnarray}
To arrive at $P^{-}$, we have used $T^{+-}$ from (\ref{H31}),
performed a partial integration, and used the equation of motion in
the light-cone gauge, $\partial_{-}^{2}A_{+}=2gJ_{-}$. To remove the
star-product in the integrand of $P^{\pm}$, we have used the
property $$\int d^{d}x f(x)\star g(x)=\int d^{d}x f(x)g(x),$$ of
noncommutative star-product, and (\ref{H21-c}) stating that in the
light-cone gauge $J_{-}$ is only a function of $x^{-}$. Thus, it
turns out that although the noncommutative nature of the theory is
fully incorporated in the current $\bar{J}\sim
\omega\star\partial_{-}\omega^{-1}$, the form of $P^{\pm}$ is
exactly the same as in commutative QED$_{2}$. We will show in the
subsequent section that this will be one of the crucial point to
show that the mass spectrum of noncommutative QED$_{2}$ is exactly
the same as in commutative Schwinger model.
\subsection{Mass spectrum of noncommutative QED$_{2}$}
\noindent As we have noted at the beginning of the previous section,
to determine the mass spectrum of noncommutative Schwinger model,
the mass eigenvalue equation
$P^{2}|\Phi\rangle=M_{}^{2}|\Phi\rangle$ is to be solved. In the
light-cone coordinates, this equation reduces to
\begin{eqnarray}\label{H32-a}
P^{+}P^{-}|\Phi\rangle=M_{}^{2}|\Phi\rangle,
\end{eqnarray}
with $P^{\pm}$ given in (\ref{H32}), which is derived in light-cone
gauge $A_{-}=0$. As for the eigenfunction $|\Phi\rangle$, we
restrict ourselves, as in \cite{armoni} for QCD$_{2}$, to the
two-current sector of the Hilbert space
\begin{eqnarray}\label{H33}
|\Phi\rangle=\int_{0}^{1}
dk~\Phi(k)\tilde{J}(-k)\tilde{J}(k-1)|0\rangle,
\end{eqnarray}
where $\tilde{J}(-k)$ and $\tilde{J}(k-1)$ play the role of creation
operators. In (\ref{H33}), $\tilde{J}(k)$ is the Fourier
transformation of $\bar{J}(x^{-})$, defined by
$\tilde{J}(k^{+})=\int\frac{dx^{-}}{\sqrt{2\pi}}e^{-ik^{+}x^{-}}\bar{J}(x^{-})$.\footnote{
To keep the notations as simple as possible, we have replaced
$k^{+}$ by $k$ in (\ref{H33}).}  Using the Fourier transform of the
current in momentum space, $\tilde{J}(k)$, the momenta $P^{\pm}$
from (\ref{H32}) are given by
\begin{eqnarray}\label{H34}
P^{+}&=&\int_{0}^{\infty} dp~\tilde{J}(-k)\tilde{J}(k),\nonumber\\
P^{-}&=&\frac{g^{2}}{2\pi}\int_{0}^{\infty}
dp~\frac{1}{p^{2}}\tilde{J}(-p)\tilde{J}(p).
\end{eqnarray}
To solve (\ref{H32-a}), with $P^{\pm}$ from (\ref{H34}) and
$|\Phi\rangle$ from (\ref{H33}), we use first the algebra of
currents in Fourier-space \cite{sadooghi-schwinger}
\begin{eqnarray}\label{H35}
[\tilde{J}(p),\tilde{J}(q)]=p~\delta(p+q).
\end{eqnarray}
Then, using $P^{\pm}|0\rangle=0$, we arrive first at the relations
\begin{eqnarray}\label{H36}
[P^{+},\tilde{J}(-k)]=2k \tilde{J}(-k),\qquad\mbox{and}\qquad
[P^{-},\tilde{J}(-k)]=\frac{g^{2}}{2\pi k}\tilde{J}(-k),
\end{eqnarray}
that lead to
\begin{eqnarray}
P^{+}|\Phi\rangle&=&2\int_{0}^{1}
dk~\Phi(k)\tilde{J}(-k)\tilde{J}(k-1)|0\rangle=2|\Phi\rangle,\label{H37}\\
P^{-}|\Phi\rangle&=&\frac{g^{2}}{2\pi}\int_{0}^{1}dk~\Phi(k)\left(\frac{1}{k}+\frac{1}{1+k}\right)\tilde{J}(-k)\tilde{J}(k-1)|0\rangle.\label{H38}
\end{eqnarray}
Here $P^{\pm}$ from (\ref{H34}) are used. Multiplying (\ref{H37})
with $M^{2}\left(P^{+}\right)^{-1}$ and using the eigenvalue
equation (\ref{H32-a}), we arrive at
\begin{eqnarray}\label{H39}
P^{-}|\Phi\rangle=\frac{M_{}^{2}}{2}|\Phi\rangle.
\end{eqnarray}
Plugging now (\ref{H38}) and (\ref{H33}) in (\ref{H39}), we get
\begin{eqnarray}\label{H40}
M_{}^{2}\Phi(k)=\frac{g^{2}}{\pi}\left(\frac{1}{k}+\frac{1}{1-k}\right)\Phi(k),
\end{eqnarray}
that leads to the mass spectrum of the theory
\begin{eqnarray}\label{H41}
M^{2}=\frac{g^{2}}{\pi}\left(\frac{1}{k}+\frac{1}{1-k}\right).
\end{eqnarray}
provided $\Phi(k)\neq 0$. This result is comparable with the mass
spectrum of QCD$_{2}$ in $N_{f}\gg N_{c}$ (see equation (35) in
\cite{armoni}). As in that case, (\ref{H41}) describes a continuum
of states with masses above $2M_{\gamma}$ in a theory whose spectrum
reduces to a single non-interacting meson with mass
$M_{\gamma}=\frac{g}{\sqrt{\pi}}$ \cite{armoni}. This supports our
claim from the previous section, that in noncommutative QED$_{2}$,
photon's mass receives no $\theta$-dependent correction.
\section{Concluding remarks} \setcounter{equation}{0}\noindent
The fact that noncommutative QED$_{2}$ is, similar to commutative
QED$_{2}$, an exactly solvable model is not \textit{a priori} clear.
In this paper, we consider only QED in two-dimensional Euclidean
space, with no possibility of going to Minkowski space. We have
shown that the photon mass in noncommutative QED$_{2}$ does not
receive any correction from the noncommutativity parameter $\theta$,
and that the spectrum of noncommutative QED$_{2}$ reduces to a
single noninteracting meson with mass
$M_{\gamma}=\frac{g}{\sqrt{\pi}}$, as in commutative Schwinger
model. To do this, in the first part of the paper, the pole of the
noncommutative photon propagator is perturbatively computed in
one-loop order. Although in noncommutative QED$_{2}$, in contrast to
commutative QED$_{2}$, three additional Feynman integrals are to be
considered, it is shown that the pole of the noncommutative photon
propagator remains unchanged at this one-loop level. This result
turns out to be (perturbatively) exact in the noncommutative
parameter $\theta$ and the coupling constant $g$, as in commutative
QED$_{2}$. It is, however, limited to the two-current sector of the
Hilbert space. This is shown in the second part of the paper, by
solving the mass eigenvalue equation of noncommutative QED$_{2}$ in
the framework of an equivalent bosonized gauge theory in
two-dimensional Euclidean space. Following closely the method used
in \cite{armoni} to determine the mass spectrum of QCD$_{2}$, we
have first derived the gauged bosonized effective action
corresponding to noncommutative QED$_{2}$, and its corresponding
energy-momentum tensor in terms of noncommutative currents. Solving
then the corresponding mass eigenvalue equation, we have determined
the mass spectrum of noncommutative QED$_{2}$, which turns out to be
exactly the same as its commutative counterpart.
\par
As we have argued in Sec. III, there are two fundamental reasons for
this  result. First, as we have seen in (\ref{H32}), in the
light-cone gauge, the noncommutative momenta of the bosonized
theory, $P^{\pm}$, turn out to have the same expressions in terms of
$\bar{J}$, as their commutative counterparts. Using these $P^{\pm}$,
we are therefore left with the same mass eigenvalue equation as in
commutative QED$_{2}$. Note that although the noncommutativity is
fully incorporated in $\bar{J}\sim
\omega\star\partial_{-}\omega^{-1}$ in terms of the star-product, in
the light-cone gauge, they depend only on one of the coordinates
$x^{-}$ [see (\ref{H21-c})]. This is the main reason why under the
integral over $x^{-}$ the star-product appearing in the integrand of
(\ref{H32}) can be removed between two $\bar{J}$'s, leaving us with
the same $P^{\pm}$ as in commutative Schwinger model. Being
independent of $x^{+}$, the Fourier transform of $\bar{J}$ depends
only on $k^{+}$. This was our motivation behind the particular
choice of (\ref{H33}) as the eigenstate $|\Phi\rangle$, which is
restricted to two-current sector of the Hilbert space as in
\cite{armoni}. As for the second reason, it lies in the fact that
the corresponding current algebra (\ref{H35}) of noncommutative
QED$_{2}$, being independent of the noncommutativity parameter
$\theta$, is the same as in commutative QED$_{2}$. This
noncommutative current algebra was particularly used in Sec. III.C
to determine certain commutation relations of $P^{\pm}$ with the
currents appearing in $|\Phi\rangle$ [see (\ref{H37}) and
(\ref{H38})]. The latter are then used to determine the mass
spectrum (\ref{H41}).
\par
It would be intriguing to study the effect of nonplanar diagrams on
noncommutative photon mass. This can be realized by bosonizing
noncommutative QED$_{2}$ with \textit{invariant} currents $j_{\mu}$
and $j_{\mu}^{5}$ from (\ref{G4-b}) and (\ref{G8}), in contrast to
what is done in this paper. What changes is, in particular, the
axial anomaly equation (\ref{G10}), corresponding to the covariant
current $J_{\mu}^{5}$. As it is shown in \cite{sadooghi2001,
armoni-2002} the axial anomaly for invariant current $j_{\mu}^{5}$
receives contributions from \textit{nonplanar} diagrams. Together
with the continuity equation, the anomaly equation (\ref{G10}) is
particularly used in this paper to determine the exact form of the
currents $J_{\pm}$ in terms of the auxiliary field $U$ in the
light-cone gauge $V=1$ [see (\ref{H4}))]. Because of the form of
nonplanar anomaly \cite{sadooghi2001,armoni-2002}, requiring an
additional IR regularization \cite{sadooghi-schwinger}, the study of
the effects of nonplanar Feynman integrals on noncommutative photon
mass is a non-trivial task and will be postponed to future
publications.
\section{Acknowledgments}
\par\noindent
The authors thank H. Partovi for collaboration in the early stage of
the work and S. Rouhani and H. Arfaei for valuable discussions. One
of the authors (M. Gh.) thanks A. Armoni and J. Sonnenschein for
email correspondences.
\begin{appendix}
\section{Vacuum polarization tensor of noncommutative Schwinger
model at one-loop level} \setcounter{equation}{0}\noindent
In this appendix, we will determine the vacuum polarization tensor
of noncommutative QED$_{2}$. This is previously done in \cite{ashok}
for general $d$-dimensional case. All propagator and vertices can
therefore be read in \cite{ashok}.\footnote{We redo the computation
here, because at some stages our results turn out to be different
from \cite{ashok}.} The Feynman diagrams corresponding to the vacuum
polarization tensor are presented in Fig. 2(a)-2(d). The
corresponding Feynman integrals are given by
\begin{eqnarray}\label{A1}
\Pi^{\mu\nu}_{\mbox{\tiny{(a)}}}(p)&=&-g^{2}\int
\frac{d^{d}k}{(2\pi)^{d}}~\frac{\mbox{tr}(\gamma^{\mu}\gamma^{\alpha}\gamma^{\nu}\gamma^{\beta})}{k^{2}(p+k)^{2}}~k_{\alpha}(p+k)_{\beta},\nonumber\\
\Pi^{\mu\nu}_{\mbox{\tiny{(b)}}}(p)&=&\frac{g^{2}N_{\mbox{\tiny{g}}}}{2}\int
\frac{d^{d}k}{(2\pi)^{d}}~\frac{1-\cos(p\wedge
k)}{k^{2}(p+k)^{2}}\left(A^{\mu\nu}+(1-\xi)B^{\mu\nu}+(1-\xi)^{2}C^{\mu\nu}
\right),\nonumber\\
\Pi^{\mu\nu}_{\mbox{\tiny{(c)}}}(p)&=&\frac{g^{2}N_{\mbox{\tiny{t}}}}{2}\int
\frac{d^{d}k}{(2\pi)^{d}}~\frac{1-\cos(p\wedge
k)}{k^{2}(p+k)^{2}}(k+p)^{2}\left((d-1)\delta^{\mu\nu}-(1-\xi)(\delta^{\mu\nu}-\frac{k^{\mu}k^{\nu}}{k^{2}})\right),\nonumber\\
\Pi^{\mu\nu}_{\mbox{\tiny{(d)}}}(p)&=&\frac{g^{2}N_{\mbox{\tiny{gh}}}}{2}\int
\frac{d^{d}k}{(2\pi)^{d}}~\frac{1-\cos(p\wedge
k)}{k^{2}(p+k)^{2}}~k^{\mu}(p+k)^{\nu},
\end{eqnarray}
where $N_{\mbox{\tiny{gh}}}=2 , N_{\mbox{\tiny{t}}}=2,
N_{\mbox{\tiny{g}}}=1$, $p\wedge k\equiv
\theta^{\mu\nu}p_{\mu}k_{\nu}$, and
\begin{eqnarray}\label{A2}
A^{\mu\nu}&=&-(5p^{2}+2k^{2}+2k\cdot p)\delta^{\mu
\nu}+(6-4d)k^{\mu}k^{\nu}+(6-d)p^{\mu}p^{\nu}+(3-2d)(k^{\mu}p^{\nu}+k^{\nu}p^{\mu}),\nonumber\\
B^{\mu\nu}&=&\left\{\frac{1}{k^{2}}\big[(k^{2}+2k\cdot
p)^{2}\delta^{\mu\nu}-\left(k^{2}+2k\cdot
p-p^{2}\right)k^{\mu}k^{\nu}- \left(k^{2}+3k\cdot
p\right)(k^{\mu}p^{\nu}+k^{\nu}p^{\mu})+p^{\mu}p^{\nu}k^{2}\big]\right\}\nonumber\\
&&+\left\{k\rightarrow
p+k,p\rightarrow -p\right\},\nonumber\\
C^{\mu\nu}&=&-\frac{\big[p^{2}k^{\mu}-(k\cdot
p)p^{\mu}\big]\big[p^{2}k^{\nu}-(k\cdot
p)p^{\nu}\big]}{k^{2}(p+k)^{2}}.
\end{eqnarray}
The first integral $\Pi^{\mu\nu}_{\mbox{\tiny{(a)}}}(p)$ does not
receive any noncommutative corrections. Evaluating the integral
using standard methods, the same result (\ref{G13}) of commutative
two-dimensional QED arises. Let us therefore look at the combination
$\tilde{\Pi}^{\mu\nu}(p)=\Pi^{\mu\nu}_{\mbox{\tiny{(b)}}}(p)+\Pi^{\mu\nu}_{\mbox{\tiny{(c)}}}(p)+\Pi^{\mu\nu}_{\mbox{\tiny{(d)}}}(p)
$, which can equivalently be given as
\begin{eqnarray}\label{A3}
\tilde{\Pi}^{\mu\nu}=\sum_{i=0}^{2}(1-\xi)^{i}
\tilde{\Pi}^{\mu\nu}_{i}.
\end{eqnarray}
In what follows, we will compute $\tilde{\Pi}^{\mu\nu}_{i}, i=0,1,2$
explicitly. Let us first look at $\tilde{\Pi}^{\mu\nu}_{0}$, which
receives contribution from diagrams (b)-(d) in Fig. 2. It is given
by
\begin{eqnarray}\label{A4}
\tilde{\Pi}^{\mu\nu}_{0}&=&\frac{g^{2}}{2}\int
\frac{d^{d}k}{(2\pi)^{d}}~\frac{1-\cos(p\wedge
k)}{k^{2}(p+k)^{2}}~{\cal{R}}^{\mu\nu},
\end{eqnarray}
where
\begin{eqnarray}\label{A5}
{\cal{R}}^{\mu\nu}&\equiv&(2d-7)p^{2}\delta^{\mu\nu}+(2d-4)k^{2}\delta^{\mu\nu}+(4d-6)(p\cdot
k)\delta^{\mu\nu}+(6-d)p^{\mu}p^{\nu}+(8-4d)k^{\mu}k^{\nu}\nonumber\\
&&+ (5-2d)p^{\mu}k^{\nu}+(3-2d)p^{\nu}k^{\mu}.
\end{eqnarray}
Using the standard Feynman parametrization method, we get
\begin{eqnarray}\label{A6}
\tilde{\Pi}^{\mu\nu}_{0}&=&\frac{g^{2}}{2}\int_{0}^{1} dx \int
_{0}^{1-x}dy \int \frac{d^{d}\ell}{(2\pi)^{d}}~\frac{1-e^{ip\wedge
\ell} }{(\ell^{2}+\Delta)^{2}}\nonumber\\
&&\times
\bigg[(2d-7)p^{2}\delta^{\mu\nu}+(2d-4)(\ell-x
p)^{2}\delta^{\mu\nu}+(4d-6)p\cdot (\ell-x
p)\delta^{\mu\nu}+(6-d)p^{\mu}p^{\nu}\nonumber\\
&&~~~+ (8-4d)(\ell^{\mu}\ell^{\nu}+x^{2}p^{\mu}p^{\nu})+
(5-2d)(\ell-xp)^{\nu}p^{\mu}+(3-2d)(\ell-xp)^{\mu}p^{\nu}\bigg],
\end{eqnarray}
where $\ell\equiv k+xp$, and $\Delta\equiv x(1-x)p^{2}$. The second
term in (\ref{A3}), $\tilde{\Pi}^{\mu\nu}_{1}$ receives
contributions from diagrams (b) and (c) in Fig. 2. It is given by
\begin{eqnarray}\label{A7}
\tilde{\Pi}^{\mu\nu}_{1}&=&\frac{g^{2}}{2}\int
\frac{d^{d}k}{(2\pi)^{d}}~\frac{1-\cos(p\wedge
k)}{k^{2}(p+k)^{2}}~{\cal{S}}^{\mu\nu},
\end{eqnarray}
where
\begin{eqnarray}\label{A8}
{\cal{S}}^{\mu\nu}&\equiv& \frac{1}{k^{2}}\bigg\{(k^{2}+2k\cdot
p)^{2}\delta^{\mu\nu}-\left(k^{2}+2k\cdot
p-p^{2}\right)k^{\mu}k^{\nu}- \left(k^{2}+3k\cdot
p\right)(k^{\mu}p^{\nu}+k^{\nu}p^{\mu})\bigg\}\nonumber\\
&&+\frac{1}{(k+p)^{2}}\bigg\{ \big[(k+p)^{2}-2p\cdot(k+p)\big]^{2}
\delta^{\mu\nu}-\big[(k+p)^{2}-2p\cdot(k+p)-p^{2}\big](k+p)^{\mu}(k+p)^{\nu}\nonumber\\
&&+\big[(k+p)^{2}-3p\cdot(k+p)\big]\big[p^{\mu}(k+p)^{\nu}+p^{\nu}(k+p)^{\mu}\big]
\bigg\}+2p^{\mu}p^{\nu}-2(k+p)^{2}\left(\delta^{\mu\nu}-\frac{k^{\mu}k^{\nu}}{k^{2}}\right).\nonumber\\
\end{eqnarray}
Introducing the Feynman parameter $x$ and after a lengthy but
straightforward calculation, (\ref{A7}) can be given by
\begin{eqnarray}\label{A9}
\tilde{\Pi}^{\mu\nu}_{1}&=&4g^{2}\int_{0 }^{1} dx(1-x)\int
\frac{d^{d}\ell}{(2\pi)^{d}}~\frac{1-e^{ip\wedge
\ell}}{(\ell^{2}+\Delta)^{3}}\nonumber\\
&&\times\bigg\{\big[(3-2x)x^{2}p^{4}-(2x+1)\ell^{2}p^{2}+4(1-x)(\ell\cdot
p)^{2}\big]\delta^{\mu\nu}+2p^{2}\ell^{\mu}\ell^{\nu}\nonumber\\
&&~~~+(2x-3)(\ell\cdot p)(\ell^{\mu} p^{\nu}+\ell^{\nu}
p^{\mu})+\big[(1+2x)\ell^{2}-(3-2x)x^{2}p^{2}\big]p^{\mu}p^{\nu}
\bigg\},
\end{eqnarray}
where $\ell=k+xp$. Similarly, $\tilde{\Pi}^{\mu\nu}_{2}$, receiving
contribution from diagram (b) in Fig. 2, is given by
\begin{eqnarray}\label{A10}
\tilde{\Pi}^{\mu\nu}_{2}&=&\frac{g^{2}}{2}\int
\frac{d^{d}k}{(2\pi)^{d}}~\frac{1-\cos(p\wedge
k)}{k^{2}(p+k)^{2}}~{\cal{T}}^{\mu\nu},
\end{eqnarray}
with
\begin{eqnarray}\label{A11}
{\cal{T}}^{\mu\nu}\equiv -\frac{\big[p^{2}k^{\mu}-(k\cdot
p)p^{\mu}\big]\big[p^{2}k^{\nu}-(k\cdot
p)p^{\nu}\big]}{k^{2}(p+k)^{2}}.
\end{eqnarray}
After some straightforward calculation, we arrive at
\begin{eqnarray}\label{A12}
\tilde{\Pi}^{\mu\nu}_{2}&=&\frac{g^{2}}{2}\int_{0}^{1} dx(1-x)\int
\frac{d^{d}\ell}{(2\pi)^{d}}~\frac{1-e^{ip\wedge
\ell}}{(\ell^{2}+\Delta)^{4}}\bigg\{(\ell^{\mu}\ell^{\nu}+x^{2}p^{\mu}p^{\nu})p^{4}\nonumber\\
&&-\big[(\ell\cdot p)\ell^{\mu}+x^{2}p^{2}p^{\mu}\big]p^{2}p^{\nu}
-\big[(\ell\cdot p)\ell^{\nu}+x^{2}p^{2}p^{\nu}\big]p^{2}p^{\mu}+
\big[(\ell\cdot p)^{2}+x^{2}p^{4}\big]p^{\mu}p^{\nu} \bigg\}.
\end{eqnarray}
To proceed, we will use the general relation
\begin{eqnarray}\label{A13}
\int \frac{d^{d}\ell}{(2\pi)^{d}}\frac{e^{ip\wedge
\ell}}{(\ell^{2}+\Delta^{2})^{n}}=\frac{2\pi^{\frac{n}{2}}}{(2\pi)^{n}}\frac{1}{\Gamma(n)}\frac{1}{(\Delta^{2})^{n-\frac{d}{2}}}
\left(\frac{|\tilde{p}|\Delta}{2}\right)^{n-\frac{d}{2}}
K_{n-\frac{d}{2}}(|\tilde{p}|\Delta),
\end{eqnarray}
and
\begin{eqnarray}\label{A14}
\int
\frac{d^{d}\ell}{(2\pi)^{d}}\frac{\ell_{\mu}\ell_{\nu}e^{ip\wedge
\ell}}{(\ell^{2}+\Delta^{2})^{n}}=F_{n}\delta_{\mu\nu}+G_{n}\frac{\tilde{p}_{\mu}\tilde{p}_{\nu}}{\tilde{p}^{2}},
\end{eqnarray}
with
\begin{eqnarray}\label{A15}
F_{n}&=&\frac{\pi^{\frac{n}{2}}}{(2\pi)^{n}}\frac{1}{\Gamma(n)}\frac{1}{(\Delta^{2})^{n-1-\frac{d}{2}}}
\left(\frac{|\tilde{p}|\Delta}{2}\right)^{n-1-\frac{d}{2}} K_{n-1-\frac{d}{2}}(|\tilde{p}|\Delta)\nonumber\\
G_{n}&=&\frac{\pi^{\frac{n}{2}}}{(2\pi)^{n}}\frac{1}{\Gamma(n)}\frac{1}{(\Delta^{2})^{n-1-\frac{d}{2}}}
\bigg[(2n-2-d)\left(\frac{|\tilde{p}|\Delta}{2}\right)^{n-1-\frac{d}{2}}
K_{n-1-\frac{d}{2}}(|\tilde{p}|\Delta)\nonumber\\
&&-2
\left(\frac{|\tilde{p}|\Delta}{2}\right)^{n-\frac{d}{2}}K_{n-\frac{d}{2}}(|\tilde{p}|\Delta)
\bigg],
\end{eqnarray}
from \cite{ashok}. Adding the corresponding contributions from
diagrams (a)-(d) of Fig. 2, and using
$A+B=\delta_{\mu\nu}\Pi^{\mu\nu}$ arising from (\ref{G18}), we
arrive at $A+B$ presented in (\ref{G20})
\begin{eqnarray}\label{A16}
A+B&=&-\mu^{2}-\frac{\mu^{2}}{4}\int_{0}^{1} dx~(x(1-x))^{-1}~
\bigg\{2(1+2x)(sK_{1}(s)-1)\nonumber\\
&&-(1-\xi)\big[6\left(2x^{2}-3x+1)(sK_{1}(s)-1\right)+\left(4x^{2}-6x+1\right)\left(s^{2}K_{2}(s)-2\right)\big]\nonumber\\
&&-\frac{1}{4}(1-\xi)^{2}(5s^{2}K_{2}(s)-2-s^{3}K_{3}(s)) \bigg\}.
\end{eqnarray}

\end{appendix}


\end{document}